\documentstyle[12pt]{article}
\title{\bf Classification of real three-dimensional Lie bialgebras and their Poisson-Lie groups}
\author{A. Rezaei-Aghdam \thanks{e-mail:rezaei-a@azaruniv.edu}, M. Hemmati and A.R. Rastkar\\{\small
Department of Physics, Azarbaijan University of Tarbiat Moallem,
53714-161, Tabriz, Iran .}}
\begin{document}
\maketitle
\begin{abstract}
Classical r-matrices of the three-dimensional real Lie bialgebras
are obtained. In this way all three-dimensional real coboundary
Lie bialgebras and their types (triangular, quasitriangular or
factorizable) are classified. Then, by using the Sklyanin bracket,
the Poisson structures on the related Poisson-Lie groups are
obtained.
\end{abstract}

\newpage

\section{\bf Introduction}
As is well known by now, the theory of classical integrable
systems is naturally related to the geometry and representation
theory of Poisson-Lie groups and the corresponding Lie bialgebras
\cite{D} and their classical r-matrices  \cite{sem}(see for
example  \cite{C.P} and  \cite{kos}). Of course recently Lie
bialgebras and their Poisson-Lie groups have application in the
theory of Poisson-Lie T-dual sigma models  \cite{klim}. Up to now
there is a detailed classification of r-matrices only for the
complex semi-simple Lie algebras  \cite{B.D}. On the other hand,
recently non-semisimple Lie algebras have important role in the
physical problems. Of course there are attempts for the
classification of low dimensional Lie bialgebras $[7-11]$. In ref
\cite{OFAR}, the classification of complex three dimensional Manin
pairs related to the complex three dimensional Lie algebras has
been performed and in this way by use of the connection between
Manin triples and the $N=2$ superconformal field theory
\cite{par}, all $N=2$ structures with $c=9$ has been classified.
In \cite{J.R} and \cite{H.S}, by use of mixed Jacobi identity for
bialgebras the authors obtain all three dimensional Lie
bialgebras. Classification of the complex and real three
dimensional Lie bialgebras have been performed in \cite{Gom} on
the same footing by using extensively the notion of twisting due
to V. G. Drinfeld \cite{D}. In this manner three dimensional real
coboundary Lie bialgebras are obtianed. In Ref \cite{Gom} the
classification of three dimensional Lie algebras of ref \cite{Pat}
has been applied. On the other hand in physical models the Bianchi
classification of three dimensional Lie algebras \cite{Shep} are
applied. In Refs \cite{J.R} and \cite{H.S} and other applications
of them \cite{von}, this classification has been applied. On the
other hand in \cite{J.R} and \cite{H.S} the type of Lie bialgebras
(coboundary or not) have not been recognized. In this paper we
perform this and classify all three dimensional real coboundary
Lie bialgebras and determine their types (triangular or
quasitriangular). Furthermore we calculate Poisson structures on
the corresponding Poisson-Lie groups. In this way, one is ready to
perform the quantization of these Lie bialgebras.

 The paper is
organized as follows. In section two, we recall some basic
definitions and propositions, then review how to obtain the three
dimensional real Lie bialgebras \cite{J.R} and \cite{H.S}. By
calculating and use of automorphism groups of Bianchi algebras we
show that these Lie bialgebras are nonisomorphic. In section
three, we determine types of 44 Lie bialgebras i.e are these
coboundary (triangular or quasitriangular) or not? We list
coboundary Lie biagebras in tables 3 and 4. We list coboundary Lie
bialgebras with coboundary duals in a separate table 4. At the end
of this section we show that these coboundary Lie bialgebras are
nonisomorphic. Finally, in section four we calculate Poisson
structures on the Poisson-Lie groups by using of the Sklyanin
bracket.

\section{\bf Three dimensional real Lie bialgebras}
Let us  recall some basic definitions and propositions \cite{D},
\cite{C.P}, \cite{kos}. Let ${\bf g}$ be a finite-dimensional Lie
algebra and ${\bf g}^\ast$ be its dual space with respect to a
non-degenerate canonical pairing $( , )$ on ${\bf
g}^\ast\times{\bf g}$.\hspace{2mm}

{\bf Definition}: A {\em Lie bialgebra} structure on a Lie algebra
${\bf g}$ is a skew-symmetric linear map $\delta : {\bf g
}\longrightarrow {\bf g}\otimes{\bf g}$ ({\em the cocommutator})
such that:

a)$\delta$ is a one-cocycle, i.e.:
\begin{equation}
\delta([X,Y])=[\delta(X),  1\otimes Y+ Y\otimes 1] + [1\otimes X+
X\otimes 1,  \delta(Y)]   \qquad \forall X,Y\in {\bf g}.
\end{equation}

b) The dual map $\delta^t:{\bf g}^\ast\otimes {\bf g}^\ast \to
{\bf g}^\ast$ is a Lie bracket on ${\bf g}^\ast$:
\begin{equation}
(\xi\otimes\eta , \delta(X)) = (\delta^t(\xi\otimes\eta) , X) =
([\xi,\eta]_\ast , X)    \qquad \forall X\in  {\bf g} ;\;\,
\xi,\eta\in{\bf g}^\ast.
\end{equation}
The Lie bialgebra defined in this way will be denoted by $({\bf
g},{\bf g}^\ast)$ or $({\bf g},\delta)$. Notice that the notation
$({\bf g},{\bf g}^\ast)$ is less precise since as we will see
there might be several nonequivalent one-cocycles on $\bf g$
giving isomorphic Lie algebra structures to ${\bf g}^\ast$,
however because of consistency and application of the results of
Refs \cite{J.R}, \cite{H.S} we will consider the notions $({\bf
g},{\bf g}^\ast)$. \hspace{2mm}

 {\bf Proposition}: One-cocycles
$\delta$ and $\delta^\prime$ of the algebra $\bf g$ are said to be
{\em equivalent} if there exists an automorphism $O$ of ${\bf g}$
such that:
\begin{equation}
\delta^\prime = (O\otimes O)\circ\delta\circ O^{-1}.
\end{equation}
In this case two Lie bialgebras $({\bf g},\delta)$ and $({\bf
g},\delta^\prime)$ are equivalent \cite{C.P},
\cite{kos}.\hspace{2mm}

{\bf Definition}:A Lie bialgebra is called {\em coboundary} Lie
bialgebra if the cocommutator is a one-coboundary, i.e, if there
exist an element $r\in{\bf g}\otimes{\bf g}$ such that:
\begin{equation}
\delta(X) = [1 \otimes X + X \otimes 1 , r]        \qquad \forall
X\in {\bf g}.
\end{equation}
\hspace{2mm}

{\bf Proposition}: Two coboundary Lie bialgebras $({\bf g},{\bf
g}^\ast)$ and $({\bf g}^\prime,{{\bf g}^\ast}^\prime)$ defined by
$r\in \bf g \otimes \bf g$ and $r^\prime \in {\bf g}^\prime
\otimes {\bf g}^\prime$ are {\em isomorphic} if and only if there
is an isomorphism of Lie algebras $\alpha: \bf g \longrightarrow
{\bf g}^\prime$ such that $(\alpha\otimes\alpha)r - r^\prime$ is
${\bf g}^\prime$ invariant i.e:
\begin{equation}
[1 \otimes X + X \otimes 1 , (\alpha\otimes\alpha)r - r^\prime]=0
\qquad \forall X\in {\bf g}^\prime.
\end{equation}
\hspace{2mm}

{\bf Definition}:Coboundary Lie bialgebras can be of two different
types:

$a$) If $r$ is a skew-symmetric solution of the classical
Yang-Baxter equation (CYBE):
\begin{equation}
[[r , r]] = 0,
\end{equation}
then the couboundary Lie bialgebra is said to be {\em triangular};
where in the above equation the schouten bracket is defined by:
\begin{equation}
[[r , r]] = [r_{12}, r_{13}] + [r_{12} , r_{23}] + [r_{13} ,
r_{23}],
\end{equation}
and if we denote $r=r^{ij}X_i \otimes X_j$, then $r_{12}=
r^{ij}X_i \otimes X_j \otimes 1$, $r_{13}= r^{ij}X_i \otimes 1
\otimes X_j$ and $r_{23}= r^{ij}1 \otimes X_i \otimes X_j$. A
solution of the CYBE is often called a {\em classical r-matrix}.

$b$) If $r$ is a solution of CYBE, such that $r_{12} + r_{21}$ is
a ${\bf g}$ invariant element of ${\bf g}\otimes{\bf g}$; then the
coboundary Lie bialgebra is said to be {\em quasi-triangular}. If
moreover, the symmetric part of $r$ is invertible, then $r$ is
called {\em factorizable}.

Sometimes condition $b$) can be replaced with the following one
\cite{D},\cite{C.P}:

$b^\prime$) If $r$ is a skew-symmetric solution of the modified
CYBE :
\begin{equation}
[[r , r]] = \omega            \qquad \omega\in {\wedge}^3 {\bf g},
\end{equation}
then the coboundary Lie bialgebra is said to be quasi-triangular.

Notice that if ${\bf g}$ is a Lie bialgebra then ${\bf g}^\ast $
is also a Lie bialgebra \cite{C.P} but this is not always true for
the coboundary property.\hspace{2mm}

 {\bf Definition}: Suppose that ${\bf g}$ be a
coboundary Lie bialgebra with one-coboundary (4); and furthermore
suppose that ${\bf g}^\ast $ is also coboundary Lie bialgebra with
the one-coboundary:
\begin{equation}
\forall\xi\in{\bf g}^\ast          \qquad\exists r^\ast\in{\bf
g}^\ast\otimes{\bf g}^\ast   \qquad \delta^{\ast}(\xi)=[1
\otimes\xi+ \xi\otimes 1 , r^\ast]_\ast,
\end{equation}
where $\delta^{\ast}: {\bf g}^\ast \longrightarrow {\bf
g}^\ast\otimes{\bf g}^\ast $. Then the pair $({\bf g},{\bf
g}^\ast)$ is called a bi-r-matrix bialgebra \cite{S}if the Lie
bracket $[ , ]^\prime$ on ${\bf g}$ defined by ${\delta^\ast}^t$:
\begin{equation}
(\delta^\ast(\xi) , X \otimes Y) = (\xi , {\delta^\ast}^t(X
\otimes Y)) = (\xi , [X,Y]^\prime)    \qquad\forall X,Y\in{\bf g}
,\qquad \xi\in{\bf g}^\ast,
\end{equation}
is equivalent to the original ones \cite{S}:
\begin{equation}
[X, Y]^\prime = S^{-1} [SX , SY]  \qquad \forall X,Y\in{\bf
g},\qquad S\in Aut({\bf g}).
\end{equation}
\hspace{2mm}

{\bf Definition} A {\em Manin} triple is a triple of Lie algebras
$(\cal{D} , {\bf g} , {\bf \tilde{g}})$ together with a
non-degenerate ad-invariant symmetric bilinear from $< , >$ on
$\cal{D}$ such that;\hspace{2mm}

 a)${\bf g}$ and ${\bf \tilde{g}}$ are Lie
subalgebras of $\cal{D}$,\hspace{2mm}

b) $\cal{D} = {\bf g}\otimes{\bf \tilde{g}}$ as a vector
space\hspace{2mm}

c) ${\bf g}$ and ${\bf \tilde{g}}$ are isotropic with respect to
$< ,>$, i.e:
\begin{equation}
<X_i , X_j> = <\tilde{X}^i , \tilde{X}^j> = 0, \hspace{16mm} <X_i
, \tilde{X}^j> = {\delta_i}^j,
\end{equation}
where $\{X_i\}$ and $\{\tilde{X}^i\}$ are the bases of the Lie
algebras ${\bf g}$ and ${\bf \tilde{g}}$, respectively. There is a
one-to-one correspondence between Lie bialgebra $({\bf g},{\bf
g}^\star)$ and Manin triple $(\cal{D} , {\bf g} , {\bf
\tilde{g}})$ with ${\bf \tilde{g}}= {\bf g}^\star$ \cite{C.P},
\cite{kos}. If we choose the structure constants of algebra ${\bf
g}$ and ${\bf \tilde{g}}$ as follows:
\begin{equation}
[X_i , X_j] = {f_{ij}}^k X_k,\hspace{20mm} [\tilde{X}^i ,\tilde{ X}^j] ={{\tilde{f}}^{ij}}_{\; \; \: k} {\tilde{X}^k}, \\
\end{equation}
then ad-invariance of the bilinear form $< , >$ on $\cal{D} = {\bf
g}\otimes{\bf \tilde{g}}$ implies that \cite{C.P}:
\begin{equation}
[X_i , \tilde{X}^j] ={\tilde{f}^{jk}}_{\; \; \; \:i} X_k +
{f_{ki}}^j \tilde{X}^k.
\end{equation}
Clearly by use of the equations (12), (13) and (2) we have:
\begin{equation}
\delta(X_i) = {\tilde{f}^{jk}}_{\; \; \; \:i} X_j \otimes X_k.
\end{equation}
By applying this relation in the one-cocycle condition (1) one can
obtain the following relation \footnote {The above relation can
also be obtained from mixed Jacobi identity for (14).}:
\begin{equation}
{f_{mk}}^i{\tilde{f}^{jm}}_{\; \; \; \; \; l} -
{f_{ml}}^i{\tilde{f}^{jm}}_{\; \; \; \; \; k} -
{f_{mk}}^j{\tilde{f}^{im}}_{\; \; \; \; \; l} +
{f_{ml}}^j{\tilde{f}^{im}}_{\; \; \; \; \; k} =
{f_{kl}}^m{\tilde{f}^{ij}}_{\; \; \; m} .
\end{equation}
In some literature the above relation is used to the definition of
Lie bialgebras.

 Now by reviewing these definitions and
propositions we are ready to review the works about three
dimensional real Lie bialgebras. In fact, in \cite{J.R} we had
applied the above relations for obtaining 28 real three
dimensional Bianchi bialgebra (Lie bialgebras where its duals are
of Bianchi type). In so doing, we had considered the Behr's
classification of three dimensional Bianchi Lie algebras
\cite{Shep}, as follows:
$$
[X_1 , X_2] = -aX_2 + n_3X_3, \hspace{20mm}[X_2 , X_3] = n_1X_1,
$$
\begin{equation}
[X_3 , X_1] = n_2X_2 + aX_3,
\end{equation}
where the structure constants are given in table 1.
\begin{center}
\hspace{3mm}{\bf Table 1} : \hspace{.5mm}Bianchi classification of
three dimensional Lie algebras.
\begin{tabular}{|c|c|c|c|c|} \hline\hline
Type                       & $a$ & $n_1$ & $n_2$ & $n_3$  \\
\hline
$I$                        & 0   & 0     & 0     & 0      \\
$II$                       & 0   & 1     & 0     & 0      \\
$VII_0$                    & 0   & 1     & 1     & 0      \\
$VI_0$                     & 0   & 1     & -1    & 0      \\
$IX$                       & 0   & 1     & 1     & 1      \\
$VIII$                     & 0   & 1     & 1     & -1     \\
$V$                        & 1   & 0     & 0     & 0      \\
$IV$                       & 1   & 0     & 0     & 1      \\
$VII_a$                    & $a$ & 0     & 1     & 1      \\
$ \left. \begin{array}{ll}
III  & (a=1) \\
VI_a & (a\neq1)
\end{array} \right\}
$
                           & $a$ & 0     & 1     & -1     \\  \hline
\end{tabular}
\end{center}

Then by considering the dual Lie algebra in the form (17) and
using the relation (16) we had obtained all Bianchi bialgebras. In
ref \cite{H.S} Hlavaty and Snobl, by considering dual algebras
which are isomorphic to Bianchi algebras ${\bf \tilde{g}}$, have
obtained (complete list) 44 real three dimensional Lie bialgebras.
These isomorphism must be such that the ad-invariant metric (12)
remains invariant under this transformations i.e:
\begin{equation}
\tilde{X}^{\prime j} = A^j_{\; \; k} \tilde{X}^k      \qquad ,
\qquad {X^\prime}_i = X_k (A^{-1})^k_{\; \; i}.
\end{equation}
Their list of 44 Lie bialgebras contain 19 Lie bialgebras of our
list in \cite{J.R}, with the names $({\bf g}, I)$,
$(VII_a,II)=(VII_a, II.i)$, $(VII_o,V)=(VII_o, V.i)$, $(VI_a,II)$,
$(VI_o,II)$, $(VI_o,V)=(VI_o, V.i)$, $(V,II)=(V, II.i)$,
$(IV,II)=(IV, II.i)$ and $(III,II)$. Notice that these 44 Lie
bialgebras are non-isomorphic. For the 28 Lie bialgebras that we
have previously obtained, it is trivial. For the other pair of Lie
bialgebras such as $({\bf g},{\bf \tilde{g}})$ and $({\bf g},{\bf
\tilde{g}}^\prime)$ where ${\bf \tilde{g}} \cong {\bf
\tilde{g}}^\prime$, as we have previously mentioned for
investigation of the Lie bialgebra isomorphism, we must examine if
relation (3) holds or not. By using (15) we can rewrite relation
(3) as follows:
\begin{equation}
O_j^{\; \; i}{{\tilde{\cal Y}}^\prime}_i = O^t {\tilde{\cal Y}}_j
O,
\end{equation}
where $({\tilde{\cal Y}}_i)^{jk}=-{\tilde{f}^{jk}}_{\; \; \; \:i}$
; $({{\tilde{\cal Y}}^\prime}_i)^{jk}=-{\tilde{f}^{\prime jk}}_{\;
\; \; \:i}$ and we apply the matrix representation of the
automorphism of the algebra ${\bf g}$ as follows:
\begin{equation}
O(X_i) = O_i^{\; \; j} X_j.
\end{equation}
In this manner for investigation of isomorphism of such Lie
bialgebras we must first obtain the automorphism groups of Bianchi
algebras.

 The automorphism groups of the complex three dimensional solvable
Lie algebras were found previously in \cite{OFAR}. Here we find
the automorphism groups of Bianchi algebras. These are Lie
subgroups of $GL(3,R)$ which preserve the Lie brackets
i.e:\footnote{Notice that these are outer automorphism groups.}
\begin{equation}
[X_i , X_j] = {f_{ij}}^k X_k,\hspace{20mm}[{X^\prime}_l , {X^\prime}_m] = {f_{lm}}^n {X^\prime}_n , \\
\end{equation}
where by applying ${X^\prime}_j= O_j^{\; \; i} X_i$ we have:
\begin{equation}
O_j^{\; \; i}O{\cal X}_i = {\cal X}_j O,
\end{equation}
or,
\begin{equation}
{\cal Y}^j O_j^{\; \; i}= O {\cal Y}^i O^t,
\end{equation}
where $({\cal X}_i)_l^{\; \;j} = -{f_{i\;l}}^j$ are the adjoint
representations of the bases of algebra $\bf g$ and as we
mentioned above $( {\cal Y}^i)_{jk} = -{f_{jk}}^i$ are the
antisymmetric matrices. Now we must first find the ${\cal X}_i$ or
${\cal Y}^i$ matrices for all Lie bialgebras. In \cite{J.R}, we
have obtained general formulas for the matrices ${\cal X}_i$ and
${\cal Y}^i$.
 Now by knowing these matrices and applying relations (22) or (23)
one can caculate general form of the elements of the automorphism
groups of the Bianchi algebras. We have found  and listed these in
table 2:
\begin{center} \hspace{10mm}{\bf Table 2} :
\hspace{3mm}Automorphism groups of the
Bianchi algebras.\\
\footnotesize{\begin{tabular}{|c|c|}  \hline\hline $\bf g$
&Automorphism group\\\hline $I$     &$GL(3,R)$ \\\hline

$II$ & $\left( \begin{array}{cc}
                      detA & 0 \\
                      v & A    \\
                     \end{array} \right)$ \qquad where $A\in GL(2,\Re), v\in\Re^2$\\\hline

$VII_0$  & $\left( \begin{array}{ccc}
                      -c & d & 0 \\
                       d & c & 0 \\
                       v^t & & 1
                     \end{array} \right)$ \qquad $c,d\in\Re$ where $c$ or $d\neq 0$ and $v\in\Re^2$ \\\hline

$VI_0$   & $\left( \begin{array}{ccc}
                      c & -d & 0 \\
                       d & -c & 0 \\
                       v^t & & 1
                     \end{array} \right)$ \qquad $c,d\in\Re$ where $c$ or $d\neq 0$ and $v\in\Re^2$ \\\hline

$IX$      &  $SO(3)$ \\\hline

$ VIII$  & $SL(2,R)$ \\\hline

$III, VI_a$& $\left(\begin{array}{ccc}
                      1 &  & v^t \\
                       0 & c & d \\
                       0 & d& c
                     \end{array} \right)$ \qquad $c,d\in\Re$ where $c$ or $d\neq 0$ and $v\in\Re^2$ \\\hline

$V$       & $\left(\begin{array}{ccc}
                      1 &  & v^t \\
                       0 &  &  \\
                       0 & & A
                     \end{array} \right)$ \qquad  where $A \in GL(2,R)$ and $v\in\Re^2$ \\\hline

$IV$ & $\left( \begin{array}{ccc}
                      1 &  & v^t \\
                       0 & c & d \\
                       0 & 0& c
                     \end{array} \right)$ \qquad $c,d\in\Re$ where $c\neq 0$ and $v\in\Re^2$ \\\hline

$VII_a$  & $\left( \begin{array}{ccc}
                      1 &  & v^t \\
                       0 & c & d \\
                       0 & -d& c
                     \end{array} \right)$ \qquad $c,d\in\Re$ where $c$ or $d\neq 0$ and $v\in\Re^2$ \\\hline
\end{tabular}}
\end{center}
Now by knowing these automorphism groups we can investigate
isomorphism of the pairs of Lie bialgebras of the form $({\bf
g},{\bf \tilde{g}})$ and $({\bf g},{\bf \tilde{g}}^\prime)$ by
using relations (19). Note that the matrices ${\tilde{\cal X}}^i$
and ${\tilde{\cal Y}}_i$ have the same form as ${\cal X}_i$ and
${\cal Y}^i$ but we must replace the set$(a, n_1, n_2, n_3)$ with $(\tilde{a}, \tilde{n}_1, \tilde{n}_2, \tilde{n}_3)$.\\
$\; \; \; \;$ These matrices can be applied for the 19 Lie
bialgebras mentioned above. For the remaining 25 Lie bialgebras
one can obtain these matrices. Note that for these Lie bialgebras
the matrices ${\cal X}_i$ and ${\cal Y}^i$ can be obtained from
equations (15) of ref \cite{J.R}; for this reason one can obtain
only the matrices ${\tilde{\cal X}}^i$ and ${\tilde{\cal Y}}_i$ of
these Lie bialgebras, for example for Lie bialgebra $( IX, V|b )$
we have:
$$
{\tilde{{\cal X}}}_1 = \left( \begin{array}{ccc}
                       0 &  0  & 0  \\
                       0 &  b  & 0  \\
                       0 &  0  & b
                    \end{array} \right), {\tilde{{\cal X}}}_2 = \left( \begin{array}{ccc}
                                                               0  & -b  & 0  \\
                                                               0  &  0  & 0  \\
                                                               0  &  0  & 0
                                                             \end{array} \right), {\tilde{{\cal X}}}_3 = \left( \begin{array}{ccc}
                                                                                                        0  &  0  & -b \\
                                                                                                        0  &  0  & 0  \\
                                                                                                        0  &  0  & 0
                                                                                                      \end{array} \right),
$$
$$
{\tilde{{\cal Y}}}_1 = \left( \begin{array}{ccc}
                       0 &  0  & 0  \\
                       0 &  0  & 0  \\
                       0 &  0  & 0
                    \end{array} \right), {\tilde{{\cal Y}}}_2 = \left( \begin{array}{ccc}
                                                                0 &  b  & 0  \\
                                                               -b &  0  & 0  \\
                                                                0 &  0  & 0
                                                             \end{array} \right), {\tilde{{\cal Y}}}_3 = \left( \begin{array}{ccc}
                                                                                                        0  &  0  & b  \\
                                                                                                        0  &  0  & 0  \\
                                                                                                        -b &  0  & 0
                                                                                                      \end{array} \right).
$$
In this manner, we investigate the isomorphicity and find that the
relations (19) do not satisfy for the pair of Lie bialgebras of
the form $({\bf g},{\bf \tilde{g}})$ and $({\bf g},{\bf
\tilde{g}}^\prime)$ mentioned in \cite{H.S}. For example, for the
Lie bialgebras $(VIII, V.i|b)$ and $(VIII, V.ii|b)$ relation (19)
for $j=1$ does not satisfy and so on ... . Now we are ready to
determine how many of the 44 real three dimensional Lie bialgebras
are coboundary?

\section{\bf Three dimensional real coboundary Lie bialgebras }

In this section we determine how many of 44 Lie bialgebras are
coboundary? Therefore, we must find $r=r^{ij}X_i \otimes X_j \in
{\bf g}\otimes{\bf g}$ such that the cocommutator of Lie
bialgebras can be written as (4). By use of (4), (13) and (15) we
have:
\begin{equation}
{\tilde{\cal Y}}_i = {{\cal X}_i}^t r + r {\cal X}_i.
\end{equation}

Now by using of (24) and form of $\cal X$ , $\cal Y$ matrices we
can find the r-matrix of the Lie bialgebras. In this manner we
determine which of the Lie bialgebras are coboundary and obtain
r-matrices. Of course we also perform this work for the dual Lie
bialgebras $(\tilde{\bf g}, {\bf g})$ by using the following
equations as the same as (24):
\begin{equation}
{\cal Y}^i = ({\tilde{\cal X}}^i)^t \tilde{r} + \tilde{r}
{\tilde{\cal X}}^i,
\end{equation}
where as above ${({\tilde{\cal X}}^i)_l}^j= -{\tilde{f}^{ij}}_{\;
\; \; \:l}$ are the adjoint representations of the bases of
algebra $\tilde{\bf g}$. The results are summerized in the
following tables 3 and 4. Notice that, we also determine the
Schouten brackets of the Lie bialgebras. In this manner the type
of Lie bialgebras(triangular or quasi-triangular) are specified
and we classify all three dimensional real coboundary Lie
bialgebras. There are two points in this tables. First, we have
listed coboundary Lie bialgebras with coboundary duals in a
separate table 4. Since such structures can be specified (up to
automorphism) by pairs of r-matrices, then it is natural to call
them bi-r-matrix bialgebras (b-r-b)\cite{S}\footnote{The most
interesting applications of b-r-b are possible in the theory of
bi-Hamiltonian dynamical systems \cite{M}. In this case, the
presence of pair of r-matrices allow us to define the pair of
dynamical systems on the space which is the space of original Lie
algebras canonically identified with its dual space \cite{sem}.}.
In \cite{S}, some examples of three dimensional b-r-b have been
given. Here we give complete list of three dimensional b-r-b.
Secondly, as it is seen we have considered skew symmetric r-matrix
solutions in tables 3 and 4. Of course there are other solutions
for some Lie bialgebras of these tables. We have listed these
solutions in other table $3^\prime$. In this table Lie bialgebras
$(III,I)$, $(VI_a,I)$ and $(VIII,V.i|b)$ are factorizable Lie
bialgebras. Other Lie bialgebras of this table are quasitriangular
such as they having r-matrix solutions with invariant symmetric
part, which for special case ($c=d=e=0$) transform to triangular
solutions of tables 3 and 4.

 Notice that in the following tables $c$, $d$ and $e$ are
arbitrary nonzero constants.
\begin{center}
\hspace{10mm}{\bf Table 3} : \hspace{3mm}Three dimensional
coboundary Lie bialgebras.
\begin{tabular}{|c|c|c|}  \hline\hline
$({\bf g}, \tilde{\bf g})$       & $r$ & $[[r , r]]$    \\\hline
$(II,I)$      &$c X_1 \wedge X_2 + d X_1 \wedge X_3$& $0$\\\hline
$(VII_o,I)$   & $cX_1 \wedge X_2$  &$0$              \\\hline
$(VII_o,V.i)$ & $X_2 \wedge X_3$ &$X_1 \wedge X_2 \wedge X_3$
\\\hline
$(VI_0,I)$    & $cX_1 \wedge X_2$ &$0$                \\\hline
$(VI_0,V.i)$  & $X_2 \wedge X_3$ &$X_1 \wedge X_2 \wedge X_3$
\\\hline
$(IX,V|b)$     &$bX_2 \wedge X_3$ &$b^2X_1 \wedge X_2 \wedge X_3$
\\ \hline
$(VIII,V.i|b)$ &$bX_2 \wedge X_3$ &$b^2X_1 \wedge X_2 \wedge X_3$
\\ \hline
$(VIII,V.ii|b)$ &$-bX_1 \wedge X_2$ &$-b^2X_1 \wedge X_2 \wedge
X_3$                                                 \\ \hline
$(VIII,V.iii)$ &$-X_1 \wedge X_2 - X_2 \wedge X_3$& $0$ \\ \hline
$(IV,II.i)$ &$- X_2 \wedge X_3$& $0$                   \\ \hline
$(IV,II.ii)$ &$\frac{1}{2}X_2 \wedge X_3$& $0$          \\\hline
$(IV.ii,VI_o)$ &$\frac{1}{2}(X_1 \wedge X_3 + X_2 \wedge X_3)$&
$0$                                                     \\\hline
$(VII_a,II.i)$ &$-\frac{1}{2a}X_2 \wedge X_3$& $0$      \\\hline
$(VII_a,II.ii)$ &$\frac{1}{2a}X_2 \wedge X_3$& $0$      \\\hline
$(III,II)$ &$-\frac{1}{2}X_2 \wedge X_3$& $0$            \\\hline
$(VI_a,II)$ &$-\frac{1}{2a}X_2 \wedge X_3$& $0$      \\\hline
\end{tabular}
\end{center}

\begin{center}
\hspace{10mm}{\bf Table 4} : \hspace{3mm}Three dimensional
bi-r-matrix
bialgebras.\\
\small{\begin{tabular}{|c|c|c||c|c|c|}  \hline\hline ${\bf g}$ &
$r$ & $[[r , r]]$ &$\tilde{\bf g}$& $\tilde{r}$ & $[[\tilde{r} ,
\tilde{r}]]$                \\ \hline
$II.i$ &$c X_1 \wedge X_2 + d X_3 \wedge X_1+X_2 \wedge X_3$&$X_1 \wedge X_2 \wedge X_3$ &$V$ & $-\frac{1}{2}X_2 \wedge X_3$&$0$  \\
\hline $VI_o$ &$c X_1 \wedge X_2 - X_2 \wedge X_3+X_3 \wedge X_1$&$0$&$V.ii$&$\frac{1}{2}(X_1 \wedge X_3 + X_2 \wedge X_3)$ & $0$                                  \\
\hline $III$ &$-\frac{1}{2}(X_1 \wedge X_2 + X_3 \wedge X_1)$&$0$&$III.ii$&$X_1 \wedge X_2 + X_3 \wedge X_1$&$0$                                       \\
\hline $III$ &$-\frac{1}{2}(X_1 \wedge X_2 + X_1 \wedge X_3)$&$0$&$III.iii$&$X_1 \wedge X_2 + X_1 \wedge X_3$&$0$                                       \\
\hline $VI_a$ &$-\frac{1}{a-1}(X_1 \wedge X_2 + X_3 \wedge X_1)$&$0$&$VI_{\frac{1}{a}}.ii$&$\frac{a-1}{2}(X_1 \wedge X_2 + X_3 \wedge X_1)$&$0$                                       \\
\hline $VI_a$ &$-\frac{1}{a+1}(X_1 \wedge X_2 + X_1 \wedge X_3)$&$0$&$VI_{\frac{1}{a}}.iii$&$\frac{a+1}{2}(X_1 \wedge X_2 + X_1 \wedge X_3)$&$0$                                       \\
\hline
\end{tabular}}
\end{center}

\newpage

\begin{center}
\hspace{10mm}{\bf Table $3^\prime$}: \hspace{3mm}Three dimensional
coboundary Lie bialgebras (other solutions).
\small{\begin{tabular}{|c|c|c|} \hline\hline $({\bf g}, \tilde{\bf
g})$       & $r$ & $[[r , r]]$
\\\hline $(III,II)$ &$cX_2\bigotimes
X_2-(c+\frac{1}{2})X_2\bigotimes X_3 -(c-\frac{1}{2})X_3\bigotimes
X_2+cX_3\bigotimes X_3$ & $0$\\\hline $(II,I)$ & $eX_1\bigotimes
X_1+cX_1 \wedge X_2+dX_1 \wedge X_3 $ &$0$\\\hline $(III,I)$
&$c(-X_2\bigotimes X_2+X_2\bigotimes X_3+X_3\bigotimes
X_2-X_3\bigotimes X_3)$ &$0$
\\\hline
$(VI_a,I)$    & $c(X_2\bigotimes X_2+X_2\bigotimes
X_3+X_3\bigotimes X_2+X_3\bigotimes X_3)$ &$0$
\\\hline $(VIII,V.i|b)$ & $bX_2 \wedge X_3\pm b(X_1\bigotimes
X_1+X_2\bigotimes X_2-X_3\bigotimes X_3)$ &$0$
\\\hline
$(VI_o,V.ii)$     &$d(X_1\bigotimes X_1-X_2\bigotimes X_2)+cX_1
\wedge X_2-X_2 \wedge X_3+X_3 \wedge X_1$ &$0$
\\ \hline
$(III,III.ii)$ &$c(X_2\bigotimes X_2-X_2\bigotimes
X_3-X_3\bigotimes X_2+X_3\bigotimes X_3)-\frac{1}{2}(X_1 \wedge
X_2+X_3 \wedge X_1)$ &$0$
\\ \hline
$(III.ii,III)$ &$cX_1\bigotimes X_1+X_1 \wedge X_2+X_3 \wedge X_1$
&$0$
\\ \hline $(III,III.iii)$ &$c(X_2\bigotimes X_2-X_2\bigotimes
X_3-X_3\bigotimes X_2+X_3\bigotimes X_3)-\frac{1}{2}(X_1 \wedge
X_2-X_3 \wedge X_1)$& $0$
\\ \hline $(III.iii,III)$ &$c(X_2\bigotimes X_2-X_2\bigotimes
X_3-X_3\bigotimes X_2+X_3\bigotimes X_3)+X_1 \wedge X_2-X_3 \wedge
X_1$& $0$
\\ \hline
\end{tabular}}
\end{center}

Notice that these coboundary Lie bialgebras are non-isomorphic. In
the previous section we mentioned to the conditions (relation (5))
under which the coboundary Lie  bialgebras are isomorphic. Here we
consider this conditions in a more exact way and not formal. By
using the matrix form of the isomorphism map $\alpha: \bf g
\longrightarrow {\bf g}^\prime$ i.e:
\begin{equation}
\alpha(X_i) = \alpha_i^{\; \; j} {X^\prime}_j,
\end{equation}
then relation (5) can be rewrite as :
\begin{equation}
{{{\cal X}^\prime}_i}^t({\alpha}^t r \alpha - r^\prime) = ({{{\cal
X}^\prime}_i}^t({\alpha}^t r \alpha - r^\prime))^t,
\end{equation}
i.e, if the above matrices are symmetric then the two coboundary
Lie bialgebras $({\bf g},\tilde{\bf g})$ and $({\bf
g}^\prime,{\tilde{\bf g}}^\prime)$ are isomorphic. Note that for
some pair of Lie bialgebras the matrix $\alpha$ is the same of the
matrix $A$ which we have previously mensioned in (18) and for some
other pairs it is the combination of two $A$ matrices. To find of
the matrices $A$ one can use the relation (18) and the following
ones:
\begin{equation}
[X_i , X_j] = {f_{ij}}^k X_k,\hspace{20mm}[{X^\prime}_l , {X^\prime}_m] = {{f^\prime}_{lm}}^n {X^\prime}_n . \\
\end{equation}
Then one finds the following equation for the  matrix $A$:
\begin{equation}
A {\tilde{\cal Y}}_j A^t = {{\tilde{\cal Y}}^\prime}_i A^i_{\; \;
j},
\end{equation}
by using this relations one can find the $A$ matrices. We perform
this works and find $A$ and then $\alpha$ matrices for the pair of
some Lie bialgebras, they are listed in appendix. By using this
matrices we have found that the matrices (27) are non-symmetric;
in other words all coboundary Lie bialgebras of tables 3 and 4 are
non-isomorphic. For example note the Lie bialgebras $(V.ii, VI_o)$
and $(V, II.i)$ then by using (3) of appendix for the matrix A one
can see that the relation (27) do not satisfy.

Notice that one can not completely compare our results with the
results of \cite{Gom}. In \cite{Gom}, the author has applied the
classification of three dimensional Lie algebras that mentioned in
\cite{Pat}. Hence our results are not completely consistent with
the results \cite{Gom}. For the algebra $SO(3)=IX$ our results are
compatible with the results of \cite{Gom}, because this Lie
algebras are the same; but for other Lie algebras, because of
isomorphicity of algebras with the Bianchi ones, the results are
not exactly the same as in \cite{Gom}.

Before beginning the next section let us discuses some about the
application of classical r-matrix in the integrable systems.
Indeed one can construct integrable systems over the vector space
${\bf g}^\ast$ related to the quasitriangular Lie bialgebras
$({\bf g},{\bf \tilde{g}})$. One can perform this by using of the
following proposition \cite{C.P}:

{\bf Proposition}: Let $H$ be a smooth function on ${\bf g}^\ast$
which is invariant under coadjoint action of ${\bf G}$ (Lie group
of $\bf g$ ) and let $r\in \bf g \otimes \bf g$ be a
skew-symmetric solution of the modified CYBE. Then, the
Hamiltonian system on ${\bf g}^\ast$ with Poisson bracket $\{ ,
\}_r$ and Hamiltonian $H$ admits a Lax pair $(L,P)$. Moreover

\begin{equation}
\{L,L\}_r = [r,L \otimes 1 + 1 \otimes L].
\end{equation}
Where  $\{ , \}_r$ is the Poisson structure related to the
following Lie bracket over $\bf g$:
\begin{equation}
[X,Y]_r = [\rho(X),Y]+[X,\rho(Y)]
\end{equation}
where $\rho:{\bf g} \rightarrow {\bf g}$ is a linear map such
that:
\begin{equation}
\rho(X_i)=\sum_j r^{ij} X_j
\end{equation}
$L:{\bf g}^\ast \rightarrow {\bf g}$ is a canonical map with
$L(\xi)=(\xi\otimes1)(t)$ where $t\in \bf g \otimes \bf g$ is the
casimir element and $P(\xi)=\rho(dH(\xi)) \qquad  \forall \xi \in
{\bf g}^\ast$.

Now by using of this proposition one can construct integrable
systems related to the three dimensional quasitriangular Lie
bialgebras. For example one can see that integrable system over
the vector space $V.i$ related to the Lie bialgebras $(VIII,
V.i|b)$ is the toda system with potential $\exp{2bq}$.

\section{\bf Calculation of Poisson structures by Sklyanin bracket}
We know that for the triangular and quasitriangular Lie bialgebras
one can obtain their corresponding Poisson-Lie groups by means of
the Sklyanin bracket provided by a given skew-symmetric r-matrix
$r=r^{ij}X_i \wedge X_j$ \cite{C.P}:
\begin{equation}
\{f_1 , f_2\} = \sum_{i,j} r^{ij}((X_i^L f_1)\, (X_j^L f_2) -
(X_i^R f_1)\, (X_j^R f_2))   \qquad \forall f_1 , f_2 \in
C^\infty(G)
\end{equation}
where $X_i^L$ and $X_i^R$ are left and right invariant vector
fields on the three dimensional related Lie group $G$. In the case
that $r$ is a solution of (CYBE), the following brackets are also
Poisson structures on the group $G$:
\begin{equation}
\{f_1 , f_2\}^L = \sum_{i,j} r^{ij}((X_i^L f_1)\, (X_j^L f_2)
\end{equation}

\begin{equation}
\{f_1 , f_2\}^R = \sum_{i,j} r^{ij}((X_i^R f_1)\, (X_j^R f_2)
\end{equation}
To calculate the left and right invariant vector fields on the
group $G$ it is enough to determine the left and right one forms.
For $g\in G$ we have:
\begin{equation}
dg g^{-1} = R^i X_i     \qquad (dg g^{-1})^i = R^i = R^i_{\;j} d
x^j,
\end{equation}

\begin{equation}
g^{-1}dg = L^i X_i     \qquad (g^{-1}dg)^i = L^i = L^i_{\;j} d
x^j,
\end{equation}
where $x^i$ are parameters of the group spaces. Now from
$\delta_j^{\;i} = <X_j^R , R^i> $ and $\delta_j^{\;i} = <X_j^L ,
L^i> $ where $X_j^R = {X^R}_j^{\;l}\partial_l $ and $X_j^L =
{X^L}_j^{\;l}\partial_l $, we obtain:
\begin{equation}
{X^R}_j^{\;l} = (R^{-t})_j^{\;\;l}   , \qquad  {X^L}_j^{\;\;l}
=(L^{-t})_j^{\;l}
\end{equation}
To calculate the above matrices we assume the following
parameterization of the group $G$:
\begin{equation}
g = e^{x_1 X_1} e^{x_2 X_2}e^{x_3 X_3}.
\end{equation}
Then, in general, for left and right invariant Lie algebra valued
one forms we have:
\begin{equation}
dg g^{-1} = dx_1 X_1 + dx_2 e^{x_1 X_1}X_2 e^{-x_1 X_1} + dx_3
e^{x_1 X_1}(e^{x_2 X_2}X_3 e^{-x_2 X_2}) e^{-x_1 X_1},
\end{equation}
\begin{equation}
g^{-1}dg  = dx_1 e^{-x_3 X_3}(e^{-x_2 X_2}X_1 e^{x_2 X_2}) e^{x_3
X_3} + dx_2 e^{-x_3 X_3}X_2 e^{x_3 X_3} + dx_3 X_3.
\end{equation}
As it is seen in the above calculations we need to calculate
expressions such as $e^{-x_i X_i}X_j e^{x_i X_i}$ \footnote{Notice
that repeated indices do not imply summation.}. Indeed in
\cite{J.R} we have shown that:
\begin{equation}
e^{-x_i X_i}X_j e^{x_i X_i} = (e^{x_i{\cal X}_i})_j^{\;\;k} X_k,
\end{equation}
where summation over index k is assumed.

For Bianchi algebras the form of matrices $e^{x_i {\cal X}_i}$ are
obtained in \cite{J.R}. For other Lie algebras which are
isomorphic to the Bianchi ones we must calculate these matrices
directly from the form of ${\cal X}_i$. We have performed these
calculations only for Lie algebras ${\bf g}$ of $({\bf
g},\tilde{\bf g})$ coboundary Lie bialgebras and then have
obtained left and right invariant vector fields as given in table
5:
\newpage
\footnotesize{\begin{center} \hspace{2mm}{\bf Table 5.1} :
\hspace{3mm}left and right invariant
vector fields over 3-dimensional coboundary Bianchi groups.\\
\begin{tabular}{|c|c|c|}  \hline\hline ${\bf
g}$&$\left(\begin{array}{c}X_1^L\\ X_2^L\\X_3^L
                     \end{array} \right)$ & $\left(\begin{array}{c}X_1^R\\
                     X_2^R\\X_3^R\end{array} \right)$\\\hline $II.i$
&$\left(\begin{array}{c}\partial_1\\-{x_3}\partial_1+\partial_2\\
\partial_3\end{array} \right)$&$\left(\begin{array}{c}\partial_1\\\partial_2\\-
{x_2}\partial_1+\partial_3\end{array} \right)$\\\hline
$VII_o$&$\left(\begin{array}{c}\cos{x_3}\partial_1+\sin{x_3}\partial_2\\-\sin{x_3}\partial_1+
\cos{x_3}\partial_2\\\partial_3\end{array}
\right)$&$\left(\begin{array}{c}\partial_1\\
\partial_2\\-{x_2}\partial_1 +{x_1}\partial_2+\partial_3\end{array} \right)$
\\\hline $VI_0$ &$\left(\begin{array}{c}\cosh{x_3}\partial_1-\sinh{x_3}
\partial_2\\-\sinh{x_3}\partial_1+\cosh{x_3}\partial_2\\ \partial_3\end{array} \right)$&
$\left(\begin{array}{c}\partial_1\\\partial_2\\-{x_2}\partial_1-{x_1}\partial_2+\partial_3\end{array}
\right)$\\\hline
$IX$&$\left(\begin{array}{c}\frac{\cos{x_3}}{\cos{x_2}}\partial_1+\sin{x_3}\partial_2-
\tan{x_2}\cos{x_3}\partial_3\\\frac{-\sin{x_3}}{\cos{x_2}}\partial_1+\cos{x_3}
\partial_2+\tan{x_2}\sin{x_3}\partial_3\\\partial_3\end{array} \right)$&
$\left(\begin{array}{c}\partial_1\\\tan{x_2}\sin{x_1}\partial_1+\cos{x_1}\partial_2-\frac{\sin{x_1}}{\cos{x_2}}\partial_3
\\-\tan{x_2}\cos{x_1}\partial_1+\sin{x_1}\partial_2+\frac{\cos{x_1}}{\cos{x_2}}\partial_3\end{array}
\right)$
\\\hline $VIII$ & $\left(\begin{array}{c}\frac{\cos{x_3}}{\cos{x_2}}\partial_1+\sin{x_3}\partial_2-
\tanh{x_2}\cos{x_3}\partial_3\\\frac{\sin{x_3}}{\cosh{x_2}}\partial_1+\cos{x_3}
\partial_2+\tanh{x_2}\sin{x_3}\partial_3\\\partial_3\end{array} \right)$ &$\left(\begin{array}{c}\partial_1\\\tanh{x_2}\sin{x_1}\partial_1+
\cosh{x_1}\partial_2+\frac{\sinh{x_1}}{\cosh{x_2}}\partial_3
\\-\tanh{x_2}\cosh{x_1}\partial_1+\sinh{x_1}\partial_2+\frac{\cosh{x_1}}{\cosh{x_2}}\partial_3\end{array}
\right)$\\\hline $V$ &$\left(\begin{array}{c}\partial_1+ x_2
\partial_2+ x_3\partial_3\\ \partial_2\\
\partial_3\end{array} \right)$ & $\left(\begin{array}{c}\partial_1 \\ e^{x_1}\partial_2\\ e^{x_1}\partial_3\end{array} \right)$\\ \hline
$V.ii$ &
$\left(\begin{array}{c}e^{x_2}\partial_1+(1-e^{x_2})\partial_2+(e^{x_2}(x_3
-1)-x_3)\partial_3\\\partial_2-x_3\partial_3\\\partial_3\end{array}
\right)$
&$\left(\begin{array}{c}\partial_1\\(1-e^{-x_1})\partial_1+e^{-x_1}\partial_2\\
e^{-x_1-x_2}\partial_3\end{array} \right)$\\\hline $IV$
&$\left(\begin{array}{c}\partial_1+ x_2
\partial_2+ (x_3-x_2)\partial_3\\\partial_2\\
\partial_3\end{array} \right)$ & $\left(\begin{array}{c}\partial_1 \\ e^{x_1}\partial_1-x_1 e^{x_1}\partial_2\\ e^{x_1}\partial_3\end{array} \right)$\\ \hline
$IV.ii$ &
$\left(\begin{array}{c}e^{x_2}\partial_1+(1-e^{x_2})\partial_2-(x_2+x_3)\partial_3\\\partial_2-x_3\partial_3\\\partial_3\end{array}
\right)$
&$\left(\begin{array}{c}\partial_1\\(1-e^{-x_1})\partial_1+e^{-x_1}\partial_2- x_1 e^{-x_1}\partial_3\\
e^{-x_1}\partial_3\end{array} \right)$\\\hline
$VII_a$&$\left(\begin{array}{c}\partial_1+ (a x_2+x_3)
\partial_2+(ax_3-x_2)\partial_3\\\partial_2\\\partial_3\end{array} \right)$ &
$\left(\begin{array}{c}\partial_1 \\ e^{a
x_1}\cos{x_1}\partial_2- e^{a x_1}\sin{x_1}\partial_3\\
e^{ax_1}\sin{x_1}\partial_2+e^{ax_1}\cos{x_1}\partial_3\end{array}\right)$\\\hline
$III$&$\left(\begin{array}{c}\partial_1+
(x_2+x_3)(\partial_2+\partial_3)\\\partial_2\\\partial_3\end{array}
\right)$&$\left(\begin{array}{c}\partial_1\\\frac{1+e^{2x_1}}{2}\partial_2+
\frac{e^{2x_1}-1}{2}\partial_3\\\frac{e^{2x_1}-1}{2}\partial_2
+\frac{1+e^{2x_1}}{2}\partial_3\end{array} \right)$\\\hline
$III.ii$&$\left(\begin{array}{c}\partial_1\\e^{-x_3}\partial_2+(e^{-x_3}-1)
\partial_3\\\partial_3\end{array}\right)$
&$\left(\begin{array}{c}\partial_1\\\partial_2\\(e^{-x_2}-1)\partial_2+e^{-x_2}\partial_3\end{array}
\right)$\\\hline
$III.iii$&$\left(\begin{array}{c}e^{-x_2-x_3}\partial_1\\\partial_2\\\partial_3\end{array}\right)$
&$\left(\begin{array}{c}\partial_1\\-x_1\partial_1+\partial_2\\-x_1\partial_1+\partial_3\end{array}
\right)$\\\hline
\end{tabular}
\end{center}}

\newpage
\footnotesize{\begin{center} \hspace{2mm}{\bf Table 5.2} :
\hspace{3mm}left and right invariant
vector fields over 3-dimensional \\coboundary Bianchi groups(continue).\\
\begin{tabular}{|c|c|c|}  \hline\hline ${\bf
g}$&$\left(\begin{array}{c}X_1^L\\ X_2^L\\X_3^L
                     \end{array} \right)$ & $\left(\begin{array}{c}X_1^R\\
                     X_2^R\\X_3^R\end{array} \right)$\\\hline  $VI_a$&$\left(\begin{array}{c}\partial_1+
(ax_2+x_3)(\partial_2+\partial_3)\\\partial_2\\\partial_3\end{array}
\right)$&$\left(\begin{array}{c}\partial_1\\e^{ax_1}(\cosh{x_1}\partial_2+\sinh{x_1}\partial_3)
\\ e^{ax_1}(\sinh{x_1}\partial_2+\cosh{x_1}\partial_3)\end{array} \right)$\\\hline
$VI_{\frac{1}{a}}.ii$&$\left(\begin{array}{c}e^{x_3-x_2}\partial_1\\e^{-\alpha
x_3}\partial_2+(e^{-\alpha
x_3}-1)\partial_3\\\partial_3\end{array}\right)$
&$\left(\begin{array}{c}\partial_1\\-x_1\partial_1+\partial_2\\x_1\partial_1
+ (e^{-\alpha x_2}-1)\partial_2+e^{-\alpha
x_2}\partial_3\end{array} \right)$ ,
$\alpha=\frac{a+1}{a-1}$\\\hline
$VI_{\frac{1}{a}}.iii$&$\left(\begin{array}{c}e^{-x_2-x_3}\partial_1\\
e^{\frac{1}{\alpha}x_3}\partial_2+(1-e^{\frac{1}{\alpha}x_3})\partial_3\\\partial_3\end{array}\right)$
&$\left(\begin{array}{c}\partial_1\\-x_1\partial_1+\partial_2\\-x_1\partial_1+
(1-e^{-\frac{1}{\alpha}x_2})\partial_2+e^{-\frac{1}{\alpha}x_2}\partial_3\end{array}
\right)$ , $\alpha=\frac{a+1}{a-1}$\\\hline
\end{tabular}
\end{center}}

Now by using these results we can calculate the Poisson structures
over the group $G$. For simplicity we can rewrite relation (33) in
the following matrix form:
\begin{equation}
\{f_1 , f_2\} = \left( \begin{array}{ccc}
                      X_1^L f_1 & X_2^L f_1 & X_3^L f_1\\
                      \end{array} \right) r \left( \begin{array}{c}
                      X_1^L f_2 \\ X_2^L f_2 \\ X_3^L f_2\\
                      \end{array} \right) - \left( \begin{array}{ccc}
                      X_1^R f_1 & X_2^R f_1 & X_3^R f_1\\
                      \end{array} \right) r \left( \begin{array}{c}
                      X_1^R f_2 \\ X_2^R f_2 \\ X_3^R f_2\\
                      \end{array} \right),
\end{equation}
and similarly we can rewrite (34) and (35).

In this manner, we calculate the fundamental Poisson brackets of
all triangular and quasitriangular Lie bialgebras. The results are
given in tables 6 and 7. Notice that for triangular Lie bialgebras
we have calculated all Poisson structures (33), (34) and (35) and
have listed in the separate table 7.
\begin{center}
\hspace{10mm}{\bf Table 6} : \hspace{3mm}Poisson
brackets related to the quasi-triangular Lie bialgebras.\\
\begin{tabular}{|c|c|c|c|}  \hline\hline
$({\bf g}, \tilde{\bf g})$  & $\{x_1,x_2\}$
&$\{x_1,x_3\}$&$\{x_2,x_3\}$      \\\hline $(II.i,V)$ &$-x_2$
&$-x_3$&$0$\\\hline $(VII_o,V.i)$ &$-x_2$ &$-\sin{x_3}$
&$\cos{x_3}-1$
\\\hline $(VI_o,v.i)$  &$-x_2$&$-\sinh{x_3}$ &$\cosh{x_3}-1$
\\\hline $(IX,V|b)$
&$-b\tan{x_2}$&$-b\frac{\sin{x_3}}{\cos{x_2}}$&$b(\cos{x_3}-\frac{1}{\cos{x_2}})$
\\\hline $(VIII,V.i|b)$ & $-b\tanh{x_2}(2{\cosh}^2{x_1}-1)$&
$b\frac{\sin{x_3}-\tanh{x_2}\sinh{2x_1}}{\cosh{x_2}}$&$b(\cos{x_3}-\frac{1}{\cosh{x_2}})$
\\\hline $(VIII,V.ii|b)$ &$b\frac{-\cos{2x_3}+\cosh{x_1}\cosh{x_2}}{\cosh{x_2}}$&
$b\frac{\sinh{x_1}-\tanh{x_2}\sin{2x_3}}{\cosh{x_2}}$
&$-b\tanh{x_2}$
\\ \hline
\end{tabular}
\end{center}

\small{\begin{center} \hspace{10mm}{\bf Table 7.1} :
\hspace{3mm}Poisson
brackets related to some triangular Lie bialgebras.\\
\begin{tabular}{|c|c|c|c|c|c|c|}  \hline\hline $({\bf g},
\tilde{\bf g})$  &
$(II,I)$&$(VII_o,I)$&$(VI_o,I)$&$(VI_o,V.ii)$&$(V,II.i)$&$(V.ii,VI_o)$\\\hline
$\{x_1,x_2\}^L$ &$c$&$c$&$c$&$c$&$0$&$0$ \\\hline $\{x_1,x_3\}^L$
&$c^\prime$&$0$&$0$&$\sinh{x_3}-\cosh{x_3}$&$0$&$\frac{e^{x_2}}{2}$\\\hline
$\{x_2,x_3\}^L$&$0$&$0$&$0$&$\sinh{x_3}-\cosh{x_3}$&
$-\frac{1}{2}$&$1-\frac{e^{x_2}}{2}$\\\hline
$\{x_1,x_2\}^R$&$c$&$c$&$c$&$c-x_2-x_1$&$0$&$0$ \\\hline
$\{x_1,x_3\}^R$&$c^\prime$&$0$&$0$&$-1$&$0$&$e^{-x_1-x_2}(1-e^{-x_1})$
\\\hline $\{x_2,x_3\}^R$&$0$&$0$&$0$&$-1$&$-\frac{e^{2x_1}}{2}$&$\frac{e^{-2x_1-x_2}}{2}$\\\hline
$\{x_1,x_2\}$&$0$&$0$&$0$&$x_2+x_1$& $0$&$0$\\\hline
 $\{x_1,x_3\}$&$0$&$0$&$0$&$\sinh{x_3}-\cosh{x_3}+1$&$0$&$\frac{e^{x_2}}{2}-e^{-x_1-x_2}(1-e^{-x_1})$ \\\hline
$\{x_2,x_3\}$&$0$&$0$&$0$&$\sinh{x_3}-\cosh{x_3}+1$&
$\frac{e^{2x_1}-1}{2}$&$1-\frac{e^{x_2}+e^{-2x_1-x_2}}{2}$\\\hline

\end{tabular}
\end{center}}
\newpage
\footnotesize{\begin{center} \hspace{10mm}{\bf Table 7.2} :
\hspace{3mm}Poisson
brackets related to triangular Lie bialgebras (continue).\\
\begin{tabular}{|c|c|c|c|c|}  \hline\hline $({\bf
g}, \tilde{\bf g})$  &
$(VIII,V.iii)$&$(IV,II.i)$&$(IV,II.ii)$&$(IV.ii,VI_o)$\\\hline
$\{x_1,x_2\}^L$ &$-\frac{\cos{2x_3}}{\cosh{x_2}}$ &$0$&$0$&$0$
\\\hline
 $\{x_1,x_3\}^L$ &$-\frac{\sin{x_3}(2\tanh{x_2}\cos{x_3}+1)}{\cosh{x_2}}$ &$0$&$0$&$\frac{e^{x_2}}{2}$\\\hline
$\{x_2,x_3\}^L$&$-\tanh{x_2}-\cos{x_3}$
&$-1$&$\frac{1}{2}$&$1-\frac{e^{x_2}}{2}$\\\hline
$\{x_1,x_2\}^R$&$-\cosh{x_1}-\tanh{x_2}\cosh{2x_1} $ &$0$&$0$&$0$
\\\hline
$\{x_1,x_3\}^R$&$-\frac{\sinh{x_1}(2\tanh{x_2}\cosh{x_1}+1)}{\cosh{x_2}}
$&$-e^{2x_1}$&$\frac{e^{x_1}}{2}$&$\frac{e^{-x_1}(2-e^{-x_1})}{2}$
\\\hline $\{x_2,x_3\}^R$&$-\frac{1}{\cosh{x_2}} $ &$x_1 e^{2x_1}$&$-\frac{x_1 e^{2x_1}}{2}$&$\frac{e^{-2x_1}}{2}$\\\hline
$\{x_1,x_2\}$&$-\frac{\cos{2x_3}}{\cosh{x_2}}+\cosh{x_1}+\tanh{x_2}\cosh{2x_1}$
&$0$&$0$&$0$ \\\hline $\{x_1,x_3\}$&$
-\frac{\sin{x_3}(2\tanh{x_2}\cos{x_3}+1)-\sinh{x_1}(2\tanh{x_2}\cosh{x_1}+1)}{\cosh{x_2}}
$&$e^{2x_1}$&$-\frac{e^{2x_1}}{2}$&$\frac{e^{x_2}+e^{-x_1}(e^{-x_1}-2)}{2}$\\\hline
$\{x_2,x_3\}$&$-\tanh{x_2}-\cos{x_3}+\frac{1}{\cosh{x_2}} $
&$-1-x_1
e^{2x_1}$&$\frac{1+x_1e^{2x_1}}{2}$&$1-\frac{e^{x_2}+e^{-2x_1}}{2}$\\\hline
\end{tabular}
\end{center}}

\small{\begin{center} \hspace{10mm}{\bf Table 7.3} :
\hspace{3mm}Poisson
brackets related to triangular Lie bialgebras (continue).\\
\begin{tabular}{|c|c|c|c|c|c|}  \hline\hline $({\bf g},
\tilde{\bf g})$  &
$(III,II)$&$(III,III.ii)$&$(III,III.iii)$&$(III.ii,III)$&$(III.iii,III)$\\\hline
$\{x_1,x_2\}^L$
&0&$-\frac{1}{2}$&$-\frac{1}{2}$&$e^{-x_3}$&$e^{-x_2-x_3}$\\\hline
$\{x_1,x_3\}^L$
&$0$&$\frac{1}{2}$&$-\frac{1}{2}$&$e^{-x_3}-2$&$e^{-x_2-x_3}$\\\hline
$\{x_2,x_3\}^L$&$-\frac{1}{2}$&$x_2+x_3$&$0$&$0$&$0$\\\hline
$\{x_1,x_2\}^R$&$0$&$-\frac{1}{2}$&$-\frac{e^{2x_1}}{2}$&$2-e^{-x_2}$&$1$\\\hline
$\{x_1,x_3\}^R$&$0$&$\frac{1}{2}$&$-\frac{e^{2x_1}}{2}$&$-e^{-x_2}$&$1$\\\hline
$\{x_2,x_3\}^R$&$-\frac{e^{2x_1}}{2}$&$0$&0&0&$0$\\\hline
$\{x_1,x_2\}$&$0$&$0$&$\frac{e^{2x_1}-1}{2}$&$e^{-x_2}+e^{-x_3}-2$&$e^{-x_2-x_3}-1$\\\hline
$\{x_1,x_3\}$&$0$&$0$&$\frac{e^{2x_1}-1}{2}$&$e^{-x_2}+e^{-x_3}-2$&$e^{-x_2-x_3}-1$\\\hline
$\{x_2,x_3\}$&$\frac{e^{2x_1}-1}{2}$&$x_2+x_3$&$0$&$0$&$0$\\\hline

\end{tabular}
\end{center}}

\small{\begin{center} \hspace{10mm}{\bf Table 7.4} :
\hspace{3mm}Poisson
brackets related to some triangular Lie bialgebras (continue).\\
\begin{tabular}{|c|c|c|c|c|}  \hline\hline $({\bf g},
\tilde{\bf g})$  &
$(VI_a,II)$&$(VI_a,VI_{\frac{1}{a}}.ii)$&$(VI_a,VI_{\frac{1}{a}}.iii)$&$(VI_{\frac{1}{a}}.ii,VI_a)$\\\hline
$\{x_1,x_2\}^L$
&$0$&$-\frac{1}{a-1}$&$-\frac{1}{a+1}$&$\frac{a-1}{2}e^{-x_2+(1-\alpha)
x_3}$\\\hline $\{x_1,x_3\}^L$&$0$&$\frac{1}{a-1}
$&$-\frac{1}{a+1}$&$\frac{a-1}{2}e^{x_3-x_2}(e^{-\alpha
x_3}-2)$\\\hline $\{x_2,x_3\}^L$&$-\frac{1}{2a}$&$\alpha(x_2+x_3)
$&$\frac{x_3-x_2}{\alpha}$&$0$\\\hline
$\{x_1,x_2\}^R$&$0$&$\frac{e^{ax_1}(\sinh{x_1}-\cosh{x_1})}{a-1}$&
$-\frac{e^{\frac{x_1}{a}}(\sinh{x_1}+\cosh{x_1})}{a+1}$&$\frac{a-1}{2}(2-e^{-\alpha
x_2})$\\\hline
$\{x_1,x_3\}^R$&$0$&$-\frac{e^{ax_1}(\sinh{x_1}-\cosh{x_1})}{a-1}
$&$-\frac{e^{\frac{x_1}{a}}(\sinh{x_1}+\cosh{x_1})}{a+1}$&$-\frac{a-1}{2}e^{-\alpha
x_2}$\\\hline
$\{x_2,x_3\}^R$&$-\frac{e^{2ax_1}}{2a}$&$0$&$0$&$0$\\\hline
$\{x_1,x_2\}$&$0$&$-\frac{1+e^{ax_1}(\sinh{x_1}-\cosh{x_1})}{a-1}
$&$\frac{-1+e^{\frac{x_1}{\alpha}}(\sinh{x_1}+\cosh{x_1})}{a+1}$&$\frac{a-1}{2}(e^{-x_2+(1-\alpha)
x_3}+e^{-\alpha x_2}-2)$\\\hline $\{x_1,x_3\}$&$0$&$
\frac{1+e^{ax_1}(\sinh{x_1}-\cosh{x_1})}{a-1}$ &$
\frac{-1+e^{\frac{x_1}{\alpha}}(\sinh{x_1}+\cosh{x_1})}{a+1}$
&$\frac{a-1}{2}(e^{x_3-x_2}(e^{-\alpha x_3}-2)+e^{-\alpha x_2})$
\\\hline
$\{x_2,x_3\}$&$\frac{e^{2ax_1}-1}{2a}$&$\alpha(x_2+x_3)
$&$\frac{x_3-x_2}{2}$&$0$
\\\hline

\end{tabular}
\end{center}}

\newpage

\small{\begin{center} \hspace{10mm}{\bf Table 7.5} :
\hspace{3mm}Poisson
brackets related to some triangular\\ Lie bialgebras (continue).\\
\begin{tabular}{|c|c|c|c|}  \hline\hline $({\bf
g},\tilde{\bf g})$
&$(VI_{\frac{1}{a}}.iii,VI_a)$&$(VII_a,II.i)$&$(VII_a,II.ii)$\\\hline
$\{x_1,x_2\}^L$
&$\frac{a+1}{2}e^{-x_2-\frac{2x_3}{a+1}}$&$0$&$0$\\\hline
$\{x_1,x_3\}^L$&$\frac{a+1}{2}(e^{-x_2-x_3}-e^{-x_2-\frac{2x_3}{a+1}})$&$0$&$0$\\\hline
$\{x_2,x_3\}^L$&$0$&$-\frac{1}{2a}$&$\frac{1}{2a}$\\\hline
$\{x_1,x_2\}^R$&$\frac{a+1}{2}(2-e^{-\frac{x_2}{\alpha}})$ &
$0$&$0$\\\hline
$\{x_1,x_3\}^R$&$\frac{a+1}{2}e^{-\frac{x_2}{\alpha}}$&$0$&$0$\\\hline
$\{x_2,x_3\}^R$&$0$&$-\frac{e^{2ax_1}}{2a}$&$\frac{e^{2ax_1}}{2a}$\\\hline
$\{x_1,x_2\}$&$\frac{a+1}{2}(e^{-x_2-\frac{2x_3}{a+1}}
+e^{\frac{-x_2}{\alpha}}-2)$&$0$&$0$\\\hline
$\{x_1,x_3\}$&$\frac{a+1}{2}(e^{-x_2-x_3}-e^{-x_2-\frac{2x_3}{a+1}}
+e^{\frac{-x_2}{\alpha}})$&$0$&$0$
\\\hline
$\{x_2,x_3\}$&$0$&$\frac{e^{2ax_1}-1}{2a}$&$-\frac{e^{2ax_1}-1}{2a}$
\\\hline

\end{tabular}
\end{center}}

 Now by knowing the Poisson structures of the Poisson-Lie groups
 one can construct dynamical systems over the symplectic leaves of
 this Poisson-Lie groups as a phase spaces. This can be down
 by using of the dressing action of ${\bf G}^\ast$ (Lie group of ${\bf
 g}^\ast$) on ${\bf G}$ which is a Poisson action whose orbits are
 exactly the symplectic leaves of ${\bf G}$ \cite{C.P},
 \cite{kos}.

\section{\bf Concluding remarks}
As mentioned above by determining the types (triangular or
quasitriangular) and obtaining r-matrices and Poisson-Lie
structures of the real three dimensional Lie bialgebras one can
construct integrable systems over the vector space ${\bf
 g}^\ast$; meanwhile one is now ready to
perform the quantization of these Lie bialgebras. Furthermore, now
one can obtain Poisson-Lie T-dual sigma models over three
dimensional triangular Lie bialgebras \cite{maj}. Notice that in
\cite{maj} only example $su(2)$ was considered. On the other hand
one can investigate integrability under Poisson-Lie T duality by
studying the Poisson-Lie T dual sigma models over three
dimensional bi-r- matrix bialgebras.

\appendix {\bf Appendix }

Here, we list $\alpha$ matrices which are applied in relations
(27).

1-For the pairs ($(IV, II.i)$ , $(IV.ii,VI_o)$) and ($(IV, II.ii)$
, $(IV.ii,IV_o)$):
$$
\alpha = A = \left( \begin{array}{ccc}
                       -1 &  0  & 0  \\
                       -1 &  1  & 0  \\
                       0 &  0  & -1
                    \end{array} \right)
$$

2-For the pair $(II, I)$ and $(II.i,V)$:
$$
\alpha = A = I
$$

3-For the pair $(V, II.i)$ and $(V.ii,VI_o)$:
$$
\alpha = A = \left( \begin{array}{ccc}
                       0 &  0  & 1  \\
                       0 &  1  & 0  \\
                       b &  0  & 0
                    \end{array} \right)
$$

4-For the pairs ($(III,III.ii)$ , $(III.ii,III)$), ($(III,
III.iii)$ , $(III.ii,III)$) and ($(III,II)$ , $(III.ii,III)$):
$$
\alpha = A = \left( \begin{array}{ccc}
                       0 &  -c  & c  \\
                       -\frac{1}{2} & d  & d+e-f  \\
                       \frac{1}{2} & e  & f
                    \end{array} \right)
$$
where $c,d,e,f \in \Re$.

5-For the pairs ($(III,III.iii)$ , $(III.iii,III)$), ($(III,
III.ii)$ , $(III.iii,III)$) and ($(III,II)$ , $(III.iii,III)$):
$$
\alpha = A = \left( \begin{array}{ccc}
                       0 &  c  & c  \\
                       \frac{1}{2} & d  & f+e-d  \\
                       \frac{1}{2} & e  & f
                    \end{array} \right)
$$
where $c,d,e,f \in \Re$.

6-For the pair
$(VI_{\frac{1}{a}}.ii,VI_a)$ and $(VI_{\frac{1}{a}}.iii,VI_a)$:
$$
\alpha = A{-1}(VI_{\frac{1}{a}} \longrightarrow
VI_{\frac{1}{a}}.ii) A(VI_{\frac{1}{a}} \longrightarrow
VI_{\frac{1}{a}}.iii)
$$
where:
$$
\alpha = A(VI_{\frac{1}{a}} \longrightarrow VI_{\frac{1}{a}}.ii) =
\left( \begin{array}{ccc}
                       0 &  c  & -c  \\
                       \frac{a}{1-a} & d  & e  \\
                       -\frac{a}{1-a} & f  & d+e-f
                    \end{array} \right)
$$
and
$$
\alpha = A(VI_{\frac{1}{a}} \longrightarrow VI_{\frac{1}{a}}.iii)
= \left( \begin{array}{ccc}
                       0 &  c^\prime  & -c^\prime  \\
                       \frac{a}{1-a} & d^\prime  & e^\prime +f^\prime -d^\prime  \\
                       \frac{a}{1+a} & f^\prime  & e^\prime
                    \end{array} \right)
$$
where $c,d,e,f \in \Re$ and similarly for prime parameter.

7-For the pair $(III.ii,III)$ and $(III.iii,III)$:
$$
\alpha = A{-1}(III \longrightarrow III.ii) A(III \longrightarrow
III.iii)
$$

\end{document}